# BCZT/LSMO/BCZT multilayer films for high temperature energy storage capacitors

Afaak LAKOUADER[1,2], Abdelilah LAHMAR[1], Špela KUNEJ[3], Daoud MEZZANE[1,2], Jamal BELHADI[2], El Hassan CHOUKRI[1], Lahoucine HAJJI[1], M'barek AMJOUD[1], Zdravko KUTNJAK[3], Igor A. LUK'YANCHUK[2,4], and Mimoun EL MARSSI[2]

[1] IMED-Lab, Cadi Ayyad University, Marrakesh, 40000, Morocco

[2] LPMC, University of Picardy Jules Verne, Amiens, 80039, France

[3] Jožef Stefan Institute, Jamova Cesta 39, 1000 Ljubljana, Slovenia

[4] Department of Building Materials, Kyiv National University of Construction and Architecture, Kyiv, 03680, Ukraine

**Abstract**

$Ba_{0.85}Ca_{0.15}Zr_{0.1}Ti_{0.9}O_3/La_{0.8}Sr_{0.2}MnO_3/Ba_{0.85}Ca_{0.15}Zr_{0.1}Ti_{0.9}O_3$ (BCZT/LSMO/BCZT) sandwich films were elaborated using the sol-gel spin coating process. The dielectric properties displayed excellent thermal stability with the temperature coefficient of capacitance, *TCC*, remaining within ± 10% between -50°C and 300°C. The high energy storage density, $W_{rec}$, of 11.8 J/cm$^3$ observed in this sandwich films, is nearly twice as high as that of the BCZT films, with an efficiency, $\eta$, of 77% under a weak electric field of 800 kV/cm. Furthermore, the stability of $W_{rec}$ and $\eta$ was observed along the studied temperature interval making them promising candidates for high-temperature energy storage capacitors.

The development of renewable energy sources has significantly increased in response to concerns about global air pollution, climate change, and energy scarcity [1]. In this context, the demand for environmentally friendly capacitors operating at high temperatures with promising and stable energy storage properties has become an urgent need to meet the requirements of modern technologies, especially in advanced electronic and cutting-edge pulsed power systems. Noting that batteries and electrochemical capacitors have limitations in meeting such demand, due to their low power density and low thermal stability. These systems also face challenges, such as electrolyte breakdown at high voltages, which restricts their suitability for high-voltage applications [2]. Therefore, dielectric capacitors offer a promising alternative owing to their high power density, fast charge-discharge capabilities, long operating life, and good chemical stability [3]. Additionally, they can be tailored to specific applications requiring operation in high temperature conditions [4], as shown in **Fig. 1(a)**. Furthermore, commercial capacitors such as XNR (N = 6, 7, 8) are no longer suitable for the increasing demands on capacitor performance. For instance, BaTiO$_3$-based X7R dielectric material exhibits high permittivity that remains constant within the temperature range of -55 °C to 125 °C, with $\Delta C/C \leq 15\%$ [5], [6] (C: capacitance). However, considerable efforts still need to be made to extend the operating temperature range of these environmentally friendly systems. Purportedly, the simultaneous inclusion of Ca and Zr in the BaTiO$_3$ (BT) matrix has received particular interest regarding to the exceptional generated properties. Indeed, the perovskite solid solution (Ba$_{1-x}$Ca$_x$)(Zr$_{1-y}$Ti$_y$)O$_3$ (BCZT) has a morphotropic phase boundary (MPB) with the coexistence of rhombohedral and tetragonal phases, which is crucial for improving physical properties that encompass electrostatic energy storage [3], [7]. Unfortunately, bulk ceramics suffer from structural and microstructural defects that minimize both energy density and breakdown strength [8]. Alternatively, thin-film ferroelectrics offer a solution to meet the demand for smaller and lighter electronic devices with maintaining their physical properties. Moreover, they afford enhanced energy density owing to their ability to maintain a high polarization and breakdown strength. Several strategies have been developed to enhance energy storage performances. One such approach is to increase breakdown strength (BDS) by incorporating a dielectric material that introduces a depolarizing field, which is decisive for obtaining a high-linearity hysteresis loop and enhancing the $W_{rec}$ [9]. Another approach is to increase polarization, particularly for low-electric-field applications, through chemical substitution or multilayers [10], [11]. It is worth mentioning that researchers have successfully developed materials with remarkable energy densities at room temperature. For example, Sun *et al.* reported that epitaxial multilayer-structured thin films of BZT/BCT deposited on Nb-doped SrTiO$_3$ substrates exhibited an ultrahigh $W_{rec}$ of 51.8 J/cm$^3$ and an excellent efficiency ($\eta$) of approximately 81.2% under a very high electric field strength of 4.5MV/cm at room temperature [12]. Furthermore, Benyoussef *et al.* showed that it is possible to tailor energy storage properties in superlattices by varying the modulation periodicity [13]. However, few studies have focused on the thermal stability of the energy density of these materials at extremely high temperatures exceeding 200°C. In the present letter, we investigate the potential of Ba$_{0.85}$Ca$_{0.15}$Zr$_{0.1}$Ti$_{0.9}$O$_3$/ La$_{0.8}$Sr$_{0.2}$MnO$_3$/ Ba$_{0.85}$Ca$_{0.15}$Zr$_{0.1}$Ti$_{0.9}$O$_3$ (BCZT/LSMO/BCZT) sandwich films as an energy storage capacitor operating under low voltages and high temperature. Both the thermal stability of the coefficient of capacitance and energy storage properties were examined over an extensive range of temperatures from -50°C to 300°C, to demonstrate their potential for specific applications under high-temperature.

The BCZT/LSMO/BCZT films were deposited on Pt/TiO$_2$/SiO$_2$/Si substrates using the sol–gel spin-coating method. The details of the preparation method of solutions and films are reported in the Supplementary Materials (SM).

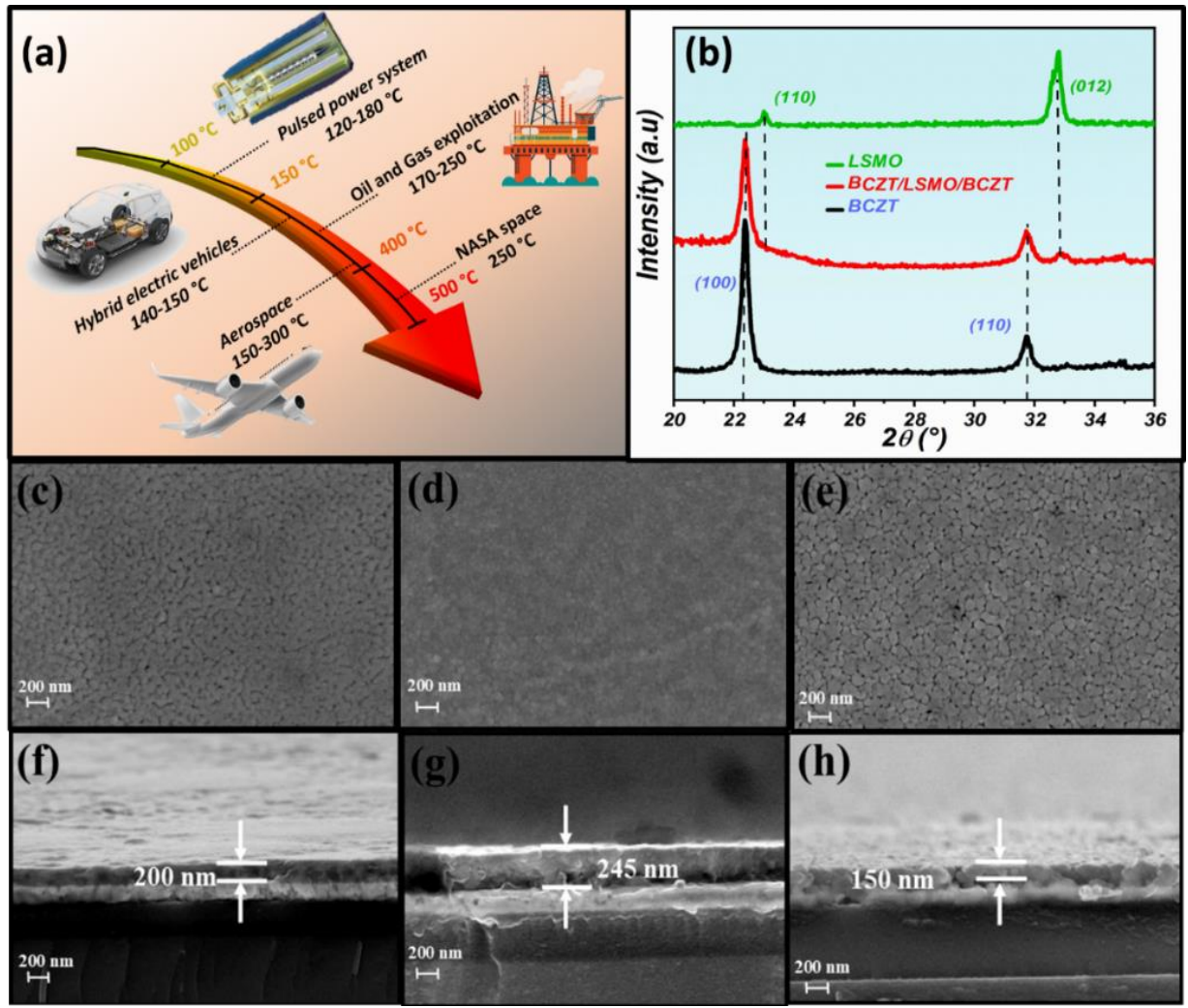

Figure 1: (a) Capacitors applications and their temperature requirements. (b) XRD patterns of BCZT, LSMO, and BCZT/LSMO/BCZT. (c)-(e) SEM images of the investigated films and (f) – (h) the corresponding cross sections, respectively.

**Fig. 1 (b)** depicts the X-ray diffraction patterns of the three investigated simple films BCZT, LSMO, and sandwich structure BCZT/LSMO/BCZT with an angular range of 20° to 36° at room temperature for better comparison. Pure perovskite phase is confirmed from the diffractograms, within the detection limit of the device. The obtained reflections are indexed based on a pseudo-cubic unit cell [7]. **Fig. 1 (c-h)** displays the top SEM images of the studied specimens. Both BCZT and LSMO thin films show a granular texture randomly distributed. However, the BCZT/LSMO/BCZT sandwich films displays a smoother surface, and a remarkably homogeneous microstructure with ultra-fine grain size distribution. The film thicknesses were approximately 200, 245, and 150 nm for BCZT, BCZT/LSMO/BCZT and LSMO, respectively, and were determined from the cross-sections as shown in **Fig. 1 (f-h).**

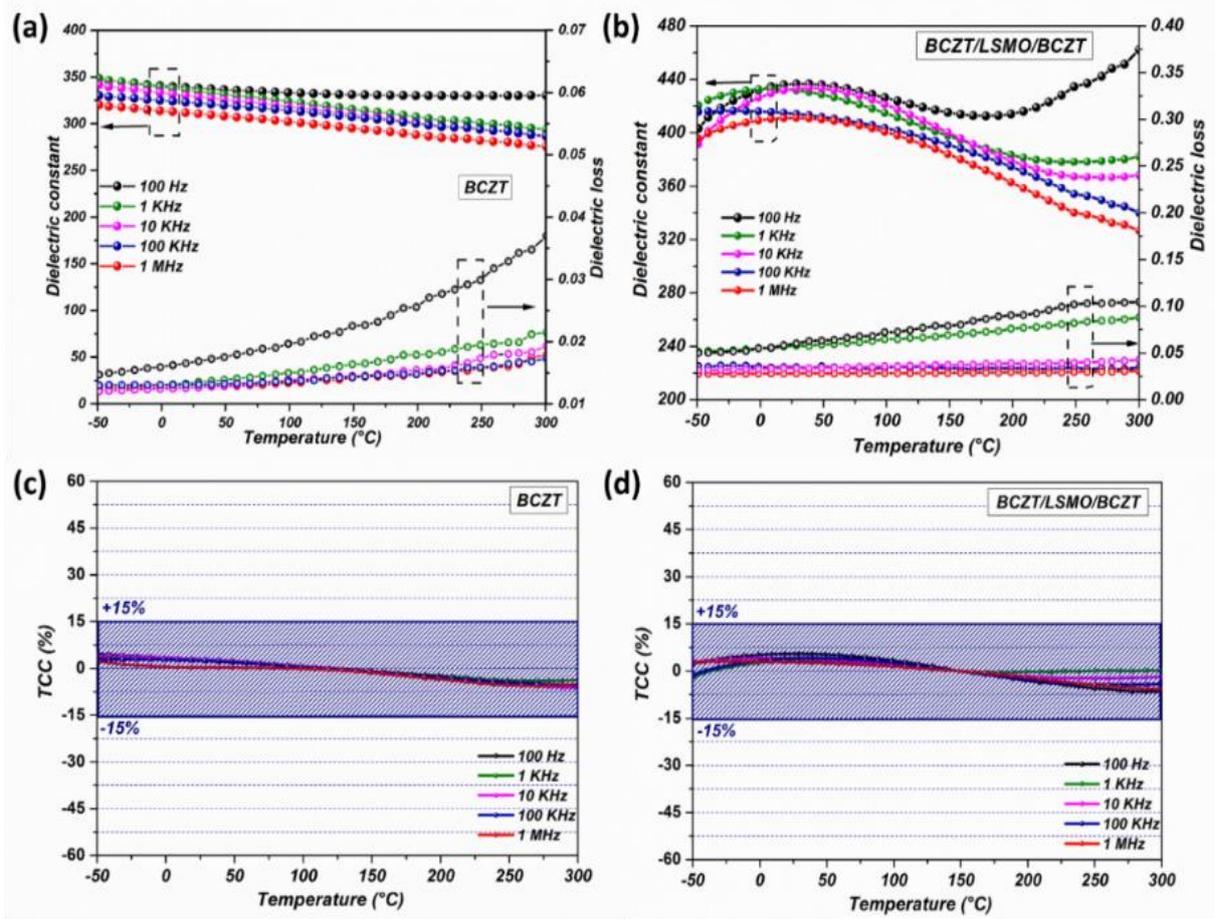

Figure 2: Thermal variation of the dielectric constant of (a) BCZT, and (b) BCZT/LSMO/BCZT. (c) and (d) show the variations of the *TCC vs* temperature of BCZT and BCZT/LSMO/BCZT, respectively.

**Fig. 2 (a)** and **(b)** depict the temperature-dependent of the dielectric constants (ε') and dielectric loss (*tan δ*) across a temperature range spanning from -50°C to 300°C at various frequencies for BCZT and BCZT/LSMO/BCZT films, respectively (**Fig. S2** displays the ε'and *tan δ* as function of frequency at room temperature, RT). The ε' of pure BCZT remains constant and slightly decreases with frequency. In fact, the increase of the frequency can limit the reorientation and movement of polarized domains, which causes a slight decrease in the dielectric constant [3]. However, BCZT/LSMO/BCZT demonstrates a maximum of ε' at low frequencies and decreased with increasing frequency. This phenomenon is commonly associated with the Maxwell-Wagner space charge polarization, which occurs at the interfaces, grain boundaries, or traps present within the various layers of the sandwich films [14]. Space charges may be formed at the interface owing to the different conductivities of BCZT and LSMO. In the LSMO phase, local charge displacement is produced by electron hopping between $Mn^{3+}$ and $Mn^{4+}$ ions [15]. This process also leads to space charge polarization and contributes to the dielectric constant at low frequency [16]. Consequently, the presence of an LSMO layer contributes to increase the dielectric constant while maintaining reduced dielectric loss. The thermal stability of the dielectric constant is established through the temperature coefficient of capacitance (*TCC*), which is calculated by comparing the dielectric constant at different temperatures to its value at a reference temperature. The *TCC* is calculated using the following equation [17]:

$$TCC(\%) = (\Delta\varepsilon_r/\varepsilon_{Base\ Temp}) \times 100 = (\varepsilon_T - \varepsilon_{Base\ Temp}/\varepsilon_{Base\ Temp}) \times 100 \qquad (1)$$

Here $\varepsilon_T$ is the dielectric constant at a certain temperature and $\varepsilon_{Base\ Temp}$ is the dielectric constant at the reference temperature, which is typically set at the midpoint of the desired operating temperature range. This temperature is often approximately 150°C, which is consistent with the value applied to high-temperature dielectric capacitors [18]. **Fig. 2(c)** and **(e)** show the variation of *TCC*, for both BCZT and BCZT/LSMO/BCZT films, across a temperature range of -50°C to-300°C. The blue striped surface, which is limited to ± 15%, represents the operational temperature range for type II capacitors [19]. Both films show good thermal stability between -50°C and 300°C. In addition, the *TCC* did not exceed a variation of ±10% across the entire frequency range explored in this study. Consequently, all samples can maintain good permittivity stability over a wide temperature range above 125°C, thereby exceeding the operational temperature range of X7R capacitors (-55°C to 125°C) [19].

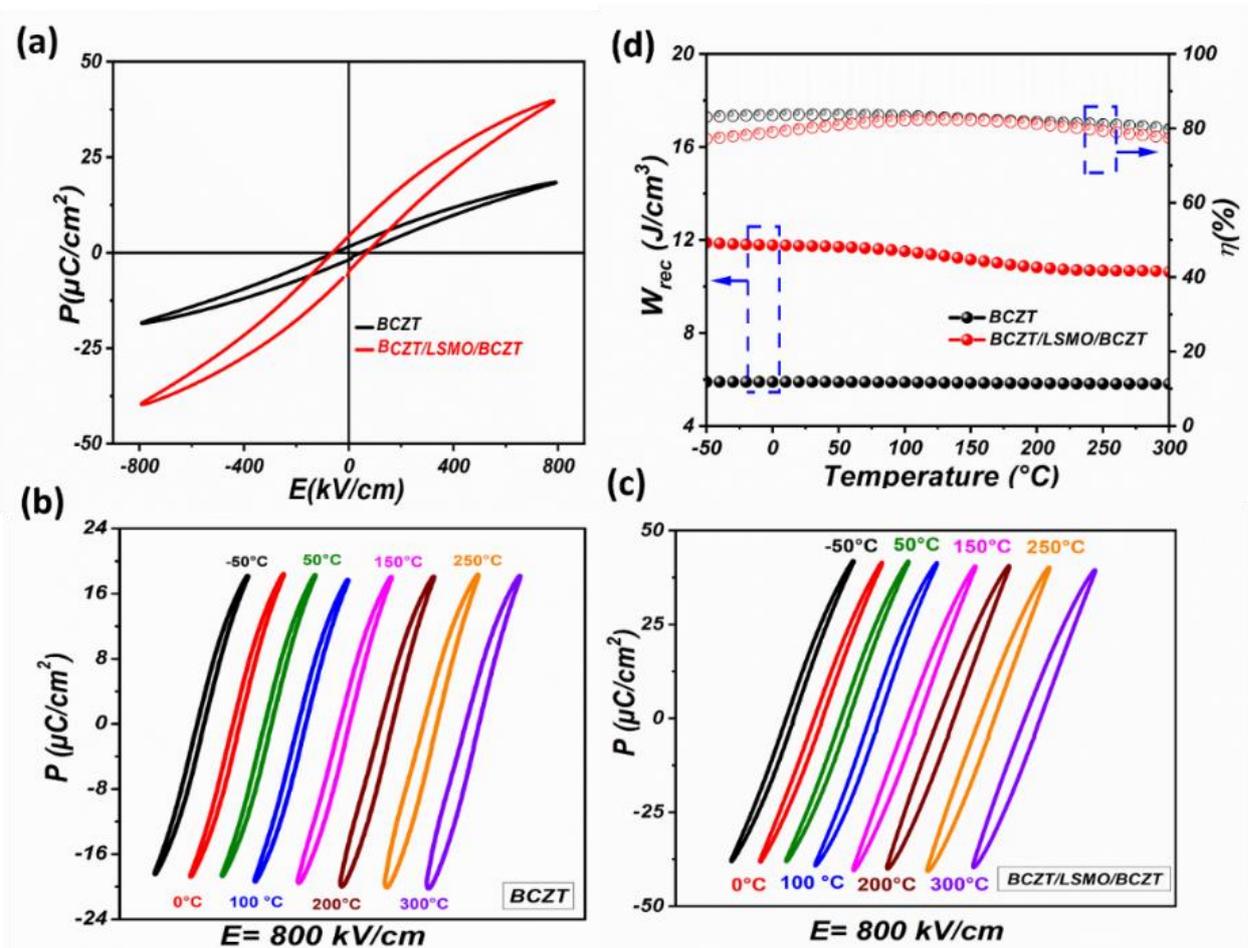

Figure 3: (a) *P-E* hysteresis loops of BCZT/LSMO/BCZT sandwich in comparison with pure BCZT thin films. The measurements were done at room temperature with an electric field of 800 kV/cm and under 10 kHz. (b) and (c) represent the temperature dependence of the *P-E* loops for BCZT and BCZT/LSMO/BCZT, respectively. (d) The variation of $W_{rec}$ and $\eta$ as a function of temperature for BCZT/LSMO/BCZT and BCZT thin films, in the temperature range of -50°C to 300°C under an electric field of 800 kV/cm.

**Fig. 3 (a)** shows the room temperature *P-E* hysteresis loops of BCZT/LSMO/BCZT and pure BCZT thin films at testing frequency of 10 kHz. A maximum polarization of $P_{max}$ = 22.5 µC/cm$^2$, and a remnant polarization of $P_r$ = 4.8 µC/cm$^2$ are obtained for BCZT thin films. Introducing LSMO as sandwiched layer (BCZT/LSMO/BCZT) leads to increase of maximum polarization to $P_{max}$ = 39.2 µC/cm$^2$, while maintaining almost the same value for the remnant one ($P_r$ = 4.7 µC/cm$^2$). This improvement could be linked to a synergistic effect of the space charge at low electric field, and also to the interlayer charge coupling, which is dominant at high field [20]. **Fig. 3 (b)** and **(c)** display the temperature-dependent *P-E* hysteresis loops for the studied films in the range of -50°C to 300°C. Both samples exhibited excellent thermal stability with maintaining elevated $P_{max}$ and low $P_r$ at high temperatures, which is essential for achieving high energy storage performance at high temperatures.

The energy storage properties were determined through the use of P-E hysteresis loops, with calculations based on the following equations [21]:

$$W_{total} = \int_0^{P_{max}} EdP \tag{2}$$

$$W_{rec} = \int_{Pr}^{P_{max}} EdP \tag{3}$$

$$\eta = \frac{W_{rec}}{W_{total}} \times 100 \tag{4}$$

Here, $W_{total}$ is the total energy, $W_{rec}$ is recoverable energy density, and $\eta$ is the energy-storage efficiency. The energy storage was calculated at RT under 1000 kV/cm as display in **Fig. S3**. The values of $W_{rec}$ and $\eta$ were 8.2 J/cm$^3$ and 72.9% for BCZT thin films and 16.3 J/cm$^3$ and 79.4% BCZT/LSMO/BCZT sandwich films, respectively. **Fig. 3(d)** despites the temperature-dependent energy storage properties at 800 kV/cm of both BCZT and BCZT/LSMO/BCZT films, over the investigated temperature range. Interestingly, the $W_{rec}$ of both films remained constant as the temperature increased. Moreover, they demonstrated high and stable efficiency ($\eta > 75\%$) across the entire temperature interval. Particularly, at 300°C the $W_{rec}$ of BCZT was 6.1 J/cm$^3$ with an efficiency of 79.8%, while BCZT/LSMO/BCZT exhibited nearly the double of energy density with $W_{rec}$ =11.8 J/cm$^3$ and related efficiency of 77%. This outcome is directly linked to the beneficial effect of the intercalation of LSMO between BCZT layers. To provide objective comparison with the literature data on equivalent materials, we used the parameter defined by Hao Pan et al.[22] as $\xi_{200°C} = W_{rec}$ (at 200°C)/$E_{app}$ (ratio of the energy storage density at 200°C to the applied electric field ($E_{app}$)), which offers trade-off between the applied field value and the overall energy storage performance. As shown in **Fig. 4**, the coefficient of $\xi_{200°C}$ for BCZT/LSMO/BCZT films is notably high and exceeds that calculated for numerous other equivalent thin films, even at a weak electric field of 800 kV/cm. This results demonstrates that this sandwich film is significant promise as advanced candidates for energy storage capacitors working at high temperature up to a 300°C. The inclusion of a single LSMO layer acts as a reagent for enhancing the energy storage performance and also ensures the preservation of desirable dielectric properties, ferroelectric properties, and high thermal stability.

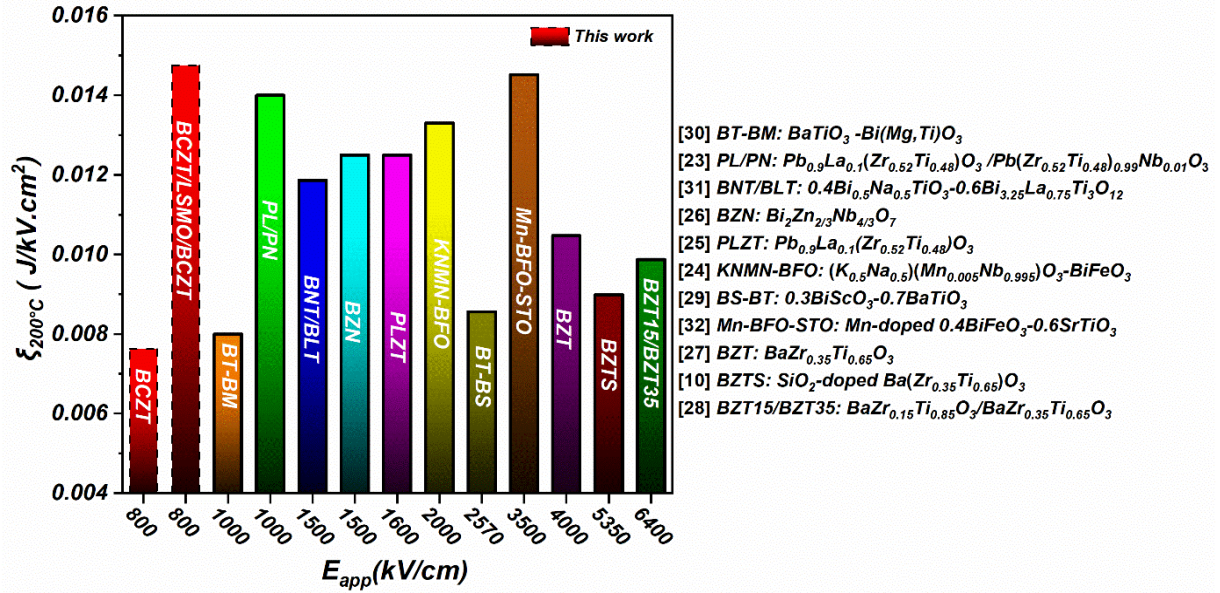

**Figure 4:** Comparison of the coefficient of $\xi_{200°C}$ for different lead-free based thin films reported in the literature [10], [23], [24], [25], [26], [27], [28], [29], [30], [31], [32].

In summary, the BCZT/LSMO/BCZT sandwich was successfully elaborated using a combined sol–gel/ spin coating. The incorporation of LSMO layer increases the dielectric constant and remains stable between -50 °C and 300 °C with a low temperature coefficient of capacitance (*TCC*) of <±10%. Moreover, a high and stable recoverable energy ($W_{rec}$ =11.8 J/cm$^3$) and efficiency ($\eta$ = of 77%) were obtained under a weak applied field of 800 kV/cm over a broad temperature range spanning from -50 °C to 300 °C. These findings underscore the superior properties of BCZT/LSMO/BCZT sandwich films as an efficient candidate as dielectric energy storage capacitors capable of meeting the urgent technological needs.

**Acknowledgments**

This work was supported by the H-GREEN project (No 101130520), and the European Union Horizon 2020 Research and Innovation Action MSCA-RISE-MELON (No. 872631). The Slovenian Research Agency grant P1-0125 is acknowledged.

**References**

[1] V. S. Puli *et al.*, "Review on energy storage in lead-free ferroelectric films," *Energy Storage*, vol. 5, no. 1, p. e359, 2023.

[2] X. Hao, "A review on the dielectric materials for high energy-storage application," *J. Adv. Dielectr.*, vol. 3, no. 01, p. 1330001, 2013.

[3] M. Maraj, W. Wei, B. Peng, and W. Sun, "Dielectric and Energy Storage Properties of Ba(1−x)CaxZryTi(1−y)O3 (BCZT): A Review," *Materials*, vol. 12, no. 21, p. 3641, Nov. 2019, doi: 10.3390/ma12213641.

[4] X. Lin, M. Salari, L. M. Reddy Arava, P. M. Ajayan, and M. W. Grinstaff, "High temperature electrical energy storage: advances, challenges, and frontiers," *Chem. Soc. Rev.*, vol. 45, no. 21, pp. 5848–5887, 2016, doi: 10.1039/C6CS00012F.

[5] A. Gurav, J. Magee, R. Phillips, and S. Carson, "C0G and X7R Ceramic Capacitors for High Temperature Applications," *Addit. Pap. Present.*, vol. 2016, no. HiTEC, pp. 000290–000298, 2016.

[6] N. Humera *et al.*, "Dielectric and ferroelectric properties of X8R perovskite barium titanate for application in multilayered ceramics capacitors," *J. Mater. Sci. Mater. Electron.*, vol. 33, no. 10, pp. 7405–7422, Apr. 2022, doi: 10.1007/s10854-022-07858-x.

[7] A. Lakouader *et al.*, "Improved energy storage and electrocaloric properties of lead-free Ba0. 85Ca0. 15Zr0. 1Ti0. 9O3 ceramic," *J. Mater. Sci. Mater. Electron.*, vol. 33, no. 18, pp. 14381–14396, 2022.

[8] A. P. Sharma, D. K. Pradhan, S. K. Pradhan, and M. Bahoura, "Large energy storage density performance of epitaxial BCT/BZT heterostructures via interface engineering," *Sci. Rep.*, vol. 9, no. 1, p. 16809, 2019.

[9] J. P. Silva *et al.*, "High-performance ferroelectric–dielectric multilayered thin films for energy storage capacitors," *Adv. Funct. Mater.*, vol. 29, no. 6, p. 1807196, 2019.

[10] F. Zhao *et al.*, "Silicon-integrated lead-free $BaTiO_3$-based film capacitors with excellent energy storage performance and highly stable irradiation resistance," *J. Mater. Chem. A*, vol. 9, no. 26, pp. 14818–14826, 2021, doi: 10.1039/D1TA03049C.

[11] C. Yang *et al.*, "Fatigue-Free and Bending-Endurable Flexible Mn-Doped Na0.5Bi0.5TiO3-BaTiO3-BiFeO3 Film Capacitor with an Ultrahigh Energy Storage Performance," *Adv. Energy Mater.*, vol. 9, no. 18, p. 1803949, 2019, doi: 10.1002/aenm.201803949.

[12] Z. Sun *et al.*, "Interface thickness optimization of lead-free oxide multilayer capacitors for high-performance energy storage," *J. Mater. Chem. A*, vol. 6, no. 4, pp. 1858–1864, 2018, doi: 10.1039/C7TA10271B.

[13] M. Benyoussef, J. Belhadi, A. Lahmar, and M. El Marssi, "Tailoring the dielectric and energy storage properties in BaTiO3/BaZrO3 superlattices," *Mater. Lett.*, vol. 234, pp. 279–282, Jan. 2019, doi: 10.1016/j.matlet.2018.09.123.

[14]   J. Rani, V. K. Kushwaha, P. K. Patel, and C. V. Tomy, "Exploring magnetoelectric coupling in trilayer [Ba(Zr0.2Ti0.8)O3- 0.5(Ba0.7Ca0.3)TiO3]/ CoFe2O4/[Ba(Zr0.2Ti0.8)O3- 0.5(Ba0.7Ca0.3)TiO3] thin film," *J. Alloys Compd.*, vol. 863, p. 157702, May 2021, doi: 10.1016/j.jallcom.2020.157702.

[15]   A. Lakouader *et al.*, "Impact of Polymeric precursor and Auto-combustion on the Structural, Microstructural, Magnetic, and Magnetocaloric Properties of La0.8Sr0.2MnO3," *J. Magn. Magn. Mater.*, vol. 586, p. 171225, Sep. 2023, doi: 10.1016/j.jmmm.2023.171225.

[16]   D. K. Pradhan *et al.*, "Exploring the Magnetoelectric Coupling at the Composite Interfaces of FE/FM/FE Heterostructures," *Sci. Rep.*, vol. 8, no. 1, p. 17381, Nov. 2018, doi: 10.1038/s41598-018-35648-1.

[17]   H. Yang, F. Yan, Y. Lin, T. Wang, and F. Wang, "High energy storage density over a broad temperature range in sodium bismuth titanate-based lead-free ceramics," *Sci. Rep.*, vol. 7, no. 1, p. 8726, 2017.

[18]   X. Liu, H. Du, X. Liu, J. Shi, and H. Fan, "Energy storage properties of BiTi0. 5Zn0. 5O3-Bi0. 5Na0. 5TiO3-BaTiO3 relaxor ferroelectrics," *Ceram. Int.*, vol. 42, no. 15, pp. 17876–17879, 2016.

[19]   A. Gurav, J. Magee, R. Phillips, and S. Carson, "C0G and X7R Ceramic Capacitors for High Temperature Applications," *Addit. Pap. Present.*, vol. 2016, no. HiTEC, pp. 000290–000298, 2016.

[20]   H. Zhu, M. Liu, Y. Zhang, Z. Yu, J. Ouyang, and W. Pan, "Increasing energy storage capabilities of space-charge dominated ferroelectric thin films using interlayer coupling," *Acta Mater.*, vol. 122, pp. 252–258, Jan. 2017, doi: 10.1016/j.actamat.2016.09.051.

[21]   S.-B. Wang *et al.*, "Large energy storage density and electrocaloric strength of Pb0.97La0.02(Zr0.46-xSn0.54Tix)O3 antiferroelectric thick film ceramics," *Scr. Mater.*, vol. 210, p. 114426, Mar. 2022, doi: 10.1016/j.scriptamat.2021.114426.

[22]   H. Pan *et al.*, "Ultrahigh energy storage in superparaelectric relaxor ferroelectrics," *Science*, vol. 374, no. 6563, pp. 100–104, Oct. 2021, doi: 10.1126/science.abi7687.

[23]   M. D. Nguyen, E. P. Houwman, M. T. Do, and G. Rijnders, "Relaxor-ferroelectric thin film heterostructure with large imprint for high energy-storage performance at low operating voltage," *Energy Storage Mater.*, vol. 25, pp. 193–201, Mar. 2020, doi: 10.1016/j.ensm.2019.10.015.

[24] S. S. Won *et al.*, "BiFeO3-doped (K0.5,Na0.5)(Mn0.005,Nb0.995)O3 ferroelectric thin film capacitors for high energy density storage applications," *Appl. Phys. Lett.*, vol. 110, no. 15, p. 152901, Apr. 2017, doi: 10.1063/1.4980113.

[25] C. T. Q. Nguyen, H. N. Vu, and M. D. Nguyen, "High-performance energy storage and breakdown strength of low-temperature laser-deposited relaxor PLZT thin films on flexible Ti-foils," *J. Alloys Compd.*, vol. 802, pp. 422–429, Sep. 2019, doi: 10.1016/j.jallcom.2019.06.205.

[26] M. Wu, S. Yu, X. Wang, and L. Li, "Ultra-high energy storage density and ultra-wide operating temperature range in Bi2Zn2/3Nb4/3O7 thin film as a novel lead-free capacitor," *J. Power Sources*, vol. 497, p. 229879, Jun. 2021, doi: 10.1016/j.jpowsour.2021.229879.

[27] "High-performance BaZr0.35Ti0.65O3 thin film capacitors with ultrahigh energy storage density and excellent thermal stability.pdf."

[28] Q. Fan *et al.*, "Significantly enhanced energy storage density with superior thermal stability by optimizing Ba(Zr0.15Ti0.85)O3/Ba(Zr0.35Ti0.65)O3 multilayer structure," *Nano Energy*, vol. 51, pp. 539–545, Sep. 2018, doi: 10.1016/j.nanoen.2018.07.007.

[29] "High energy storage efficiency and thermal stability of A-sitedeficient and 110-textured BaTiO3–BiScO3 thin films.pdf."

[30] H. Yan, B. Song, K. Zhu, L. Xu, B. Shen, and J. Zhai, "Achieved high energy density and excellent thermal stability in (1−x)(Bi0.5Na0.5)0.94Ba0.06TiO3−xBi(Mg0.5Ti0.5)O3 relaxor ferroelectric thin films," *J. Mater. Sci. Mater. Electron.*, vol. 32, no. 12, pp. 16269–16278, Jun. 2021, doi: 10.1007/s10854-021-06174-0.

[31] W. Yue *et al.*, "Bi0.5Na0.5TiO3-Bi3.25La0.75Ti3O12 Lead-Free Thin Films for Energy Storage Applications through Nanodomain Design," *Crystals*, vol. 12, no. 11, Art. no. 11, Nov. 2022, doi: 10.3390/cryst12111524.

[32] B. Wang *et al.*, "Excellent energy storage performance of Mn-doped SrTiO3-BiFeO3 thin films by microstructure modulation," *J. Alloys Compd.*, vol. 968, p. 171756, Dec. 2023, doi: 10.1016/j.jallcom.2023.171756.